\let\texyear\year
\documentclass{ieeeaccess}
\usepackage{cite}
\usepackage{amsmath,amssymb,amsfonts}
\usepackage{algorithmic}
\usepackage{graphicx}
\usepackage{textcomp}
\let\ieeeaccessyear\year
\let\year\texyear
\usepackage{caption}
\usepackage{dblfloatfix}
\usepackage{float}
\usepackage{makecell}
\usepackage{siunitx}
\usepackage{tabularx,booktabs} 
\usepackage{rotating}
\usepackage{pgf}
\let\year\ieeeaccessyear
\definecolor{accessblue}{RGB}{0,105,154}
\def\BibTeX{{\rm B\kern-.05em{\sc i\kern-.025em b}\kern-.08em
    T\kern-.1667em\lower.7ex\hbox{E}\kern-.125emX}}
\begin{document}
\history{Date of publication xxxx 00, 0000, date of current version xxxx 00, 0000.}
\doi{10.1109/ACCESS.2017.DOI}

\title{Universal Joint Feature Extraction for P300 EEG Classification using Multi-task Autoencoder}
\author{Apiwat Ditthapron\authorrefmark{1}, Nannapas Banluesombatkul\authorrefmark{2}, Sombat Ketrat\authorrefmark{2}, \\Ekapol Chuangsuwanich\authorrefmark{3} and Theerawit Wilaiprasitporn\authorrefmark{2}, \IEEEmembership{Member, IEEE}}
\address[1]{Computer Department, Worcester Polytechnic Institute, Worcester, MA, USA (e-mail: aditthapron@wpi.edu}
\address[2]{Bio-inspired Robotics and Neural Engineering Lab, School of Information Science and Technology, Vidyasirimedhi Institute of Science \& Technology, Rayong, Thailand (e-mail: theerawit.w@vistec.ac.th).}
\address[3]{Computer Engineering Department, Chulalongkorn University, Bangkok, Thailand.}
\tfootnote{This work was supported by The Thailand Research Fund and Office of the Higher Education Commission under Grant MRG6180028.}

\begin{abstract}
The process of recording Electroencephalography (EEG) signals is onerous and requires massive storage to store signals at an applicable frequency rate. In this work, we propose the Event-Related Potential Encoder Network (ERPENet); a multi-task autoencoder-based model, that can be applied to any ERP-related tasks. The strength of ERPENet lies in its capability to handle various kinds of ERP datasets and its robustness across multiple recording setups, enabling joint training across datasets. ERPENet incorporates Convolutional Neural Networks (CNNs) and Long Short-Term Memory (LSTM), in an autoencoder setup, which tries to simultaneously compress the input EEG signal and extract related P300 features into a latent vector. Here, we can infer the process for generating the latent vector as universal joint feature extraction. The network also includes a classification part for attended and unattended events classification as an auxiliary task. We experimented on six different P300 datasets. The results show that the latent vector exhibits better compression capability than the previous state-of-the-art semi-supervised autoencoder model. For attended and unattended events classification, pre-trained weights are adopted as initial weights and tested on unseen P300 datasets to evaluate the adaptability of the model, which shortens the training process as compared to using random Xavier weight initialization. At the compression rate of 6.84, the classification accuracy outperforms conventional P300 classification models: XdawnLDA, DeepConvNet, and EEGNet achieving 79.37\% - 88.52\% classification accuracy depending on the dataset.
\end{abstract}


\titlepgskip=-15pt

\maketitle

\section{Introduction}
\label{sec:introduction}
\PARstart{B}{rain} informatics-based large-scale architecture and its applications have been introduced in recent years \cite{Zhong2015}. The pipeline of brain informatics comprises data acquisition \textit{(physical layer)}, brain data center \textit{(storage and computational layer)} and service objects \textit{(application layer)}. The tools used to acquire data on the physical layer are wearable devices, non-contact sensing, high cognitive measurements (such as brain activity) and questionnaires. Then, the data is transferred to perform data management, mining, clustering, and computing (including machine learning and deep learning algorithms) in the storage and computational layer. At the end of the pipeline, the application layer provides a service for various groups of users such as researchers, physicians, healthcare-related businesses, etc. In this study, we focus on the large-scale non-invasive brain signal data called electroencephalogaphy (EEG), which is stored inside the computational layer. Regarding high sampling rate of EEG signal over multiple channels, a massive storage is required to maintain data in the size that can be used for future applications. A large volume of data has been exploited from EEG-related research such as brain-computer interface (BCI) and cognitive neuroscience to create a downstream model that capable of encoding EEG in smallest size that maintains features requires for various classification tasks in application layer.

Due to the potentially large volume of data, the application requires an effective way to compress and store the collected data. Several studies have attempted to reduce the amount of data by decreasing the length of each sample. One popular procedure achieving this is compressed sensing (CS). It projects a portion of the input signal onto a random matrix like Gaussian, sparse binary, or binomial such that the size of the projected data is smaller than the length of original samples. Prior to data analysis, we needed to reconstruct the projected data using certain variants of CS techniques such as sparse Bayesian learning \cite{zhang2013compressed, zhang2013compressed2, zhang2013extension, zhang2014spatiotemporal}, or reconstruction-based inter-channel and intra-channel correlations \cite{majumdar2014low, majumdar2014non, shukla2015row, majumdar2015energy, shukla2015exploiting}. However, the reconstruction of CS involves the solution of optimization problems which can be time-consuming, so is impractical for use in real-time applications such as online BCI for electrical appliances controls \cite{hoffmann2008efficient}.

Recently, a CS method based on deep learning (DL), namely the Autoencoder (AE), was applied in body area networks and tele-monitoring systems\cite{gogna2017}. The authors reported the advantages of AE over the classical CS technique in both computation time and accuracy for bio-signal data reconstruction. The goal of their work was not only to find the optimal data compression procedure but also to classify the event type. To combine the data compression and address classification problems, a multi-task AE was applied due to its capability of class labeling, along with reconstruction (The concept of a multi-task AE is described further in the methodology section of this paper). Epileptic, eyes-closed and eyes-opened EEGs from five subjects and five classes in total, \cite{altunay2010}, were used in their proposed model evaluation. Unlike their model which was merely formed by two fully connected (FC) layers, we proposed our own multi-task AE, using stacks of two-dimensional convolutional neural networks (2D-CNNs) and long short-term memory units (LSTMs) in order to capture both spatial and temporal information. The CNN-LSTM combined model was first introduced in the video classification problem domain which outperformed conventional spatial-temporal classification models, such as Support Vector Machine, standalone 1D-CNNs, and standalone LSTMs\cite{wu2015modeling}. Recently, the combination of CNN-LSTM has been adapted, using 2D-CNNs to extract spatial feature over EEG montage instead of video-frame feature extraction, to classify EEG signal in various tasks\cite{tan2017multimodal},\cite{wilaiprasitporn2018affective}. However, training CNN-LSTM architecture needs a sufficiently large amount of EEG data which is inconceivable in some of EEG classification tasks because of the complexity and the expense of EEG signals recording. In \cite{tan2017multimodal}, the authors proposed a multimodal classification of EEG video and optical flow to solve insufficient data problem, but the training duration was significantly longer than other methods. Here, we proposed an ERPENet, a CNN-LSTM pretrained model, to migrate insufficient data problem.

To prove the concept of brain informatics on large-scale EEG data, EEG responses known as event-related potentials (ERPs) are examined. ERPs are widely known EEG responses from brain activity related to human perceptual and cognitive processes \cite{luck2014,picton1992}. Furthermore, the major ERP components are narrowed down, namely \textit{P300} using simple experimental tasks or paradigms across BCIs and cognitive neurosciences (the oddball paradigm). A fundamental experimental design is then adopted to study P300 responses from both attended and unattended events according to the human perceptual and cognitive processes \cite{squires1975}. The attended event is one where a human is waiting to perceive the target information with a low probability of occurrence. In contrast, an unattended event perceives non-targeted information with a higher probability of appearances.  Here, six P300 datasets (large-scale EEG data) are examined from various studied events including Documenting, Modelling and Exploiting P300 Amplitude in Donchin's Speller \cite{citi2010}, BCI Competition III - Dataset II \cite{rakotomamonjy2008bci}, Auditory Multi-Class BCI \cite{schreuder2010}, BCI-Spelling using Rapid Serial Visual Presentation (RSVP) \cite{acqualagna2013}, Examining EEG-Alcoholism Correlation (control group) \cite{UCI} and Decoding Auditory Attention \cite{bnci15}behaving (more details in Section IV). These six ERPs were designed and recorded for distinct tasks, but can be classified as attended and unattended events, according to P300 classification systems. The aggregated P300 from multiple datasets was used to train a universal feature extractor as a pre-trained model, in which its pre-trained weights are able to speed up the training process and reduce the overfitting problem in any P300 models with limited training data.

The previous multi-task AE work\cite{gogna2017} also lacks the ability to handle and exploit information shared across different recording setups, whereas the proposed multi-task AE in this study is trained and validated across six datasets, obtained from various experimental studies with different numbers of EEG channels. To deal with the dataset inconsistencies, 2D-CNNs are introduced to the model, which will be further discussed in Section III. However, all have common EEG features such as P300 or ERPs from either the attended or unattended events. We conducted two experiments to verify the academic merits and novelty of our work. Firstly, an experiment was conducted to demonstrate the performance of large-scale EEG compression on an unseen P300 dataset using the purposed multi-task AE, containing ERPENet. Then, the ERPENet was adopted as a pre-trained network to the attended and unattended event classification network. The results are compared to the state-of-the-art P300 dimensionality reduction algorithm\cite{lotte2018review} named Xdawn with Bayesian LDA classification \cite{mika1999fisher} and state-of-the-art in EEG classification deep learning models, EEGNet\cite{Waytowich2018} and DeepConvNet\cite{hbm23730}. EEGNet and DeepConvNet were deep learning models, designed for various EEG classification tasks, and yielded state-of-the-art results.

Three main contributions of this work can be summarized as
follows:
\begin{itemize}
    \item A CS method that compresses P300 for classification and restoration in the application layer of Brain information-based large-scale architecture.  
    \item A robust multi-task autoencoder, composes of 2D-CNNs and LSTMs, is proposed to extract P300 features in classification task and compress EEG signal across various experimental studies with different setups of EEG recording. 
    \item ERPENet, a pre-trained encoder network of proposed multi-task autoencoder, is capable of can be fine-tuned and applied to a new related application with limited training data.
\end{itemize}

The rest of this paper is organized as follows. Section II provides backgrounds in DL which are the basis of our proposed model. Section III illustrates the designing of multi-task autoencoder in detail. Section IV presents the datasets used in the experimental studies. Section V discusses the experimental protocols used to evaluate the proposed model. Finally, the results, discussion and conclusion are contained in Sections V, VI and VII, respectively.

\begin{figure*}
 \centering
  \includegraphics[width=0.8\linewidth]{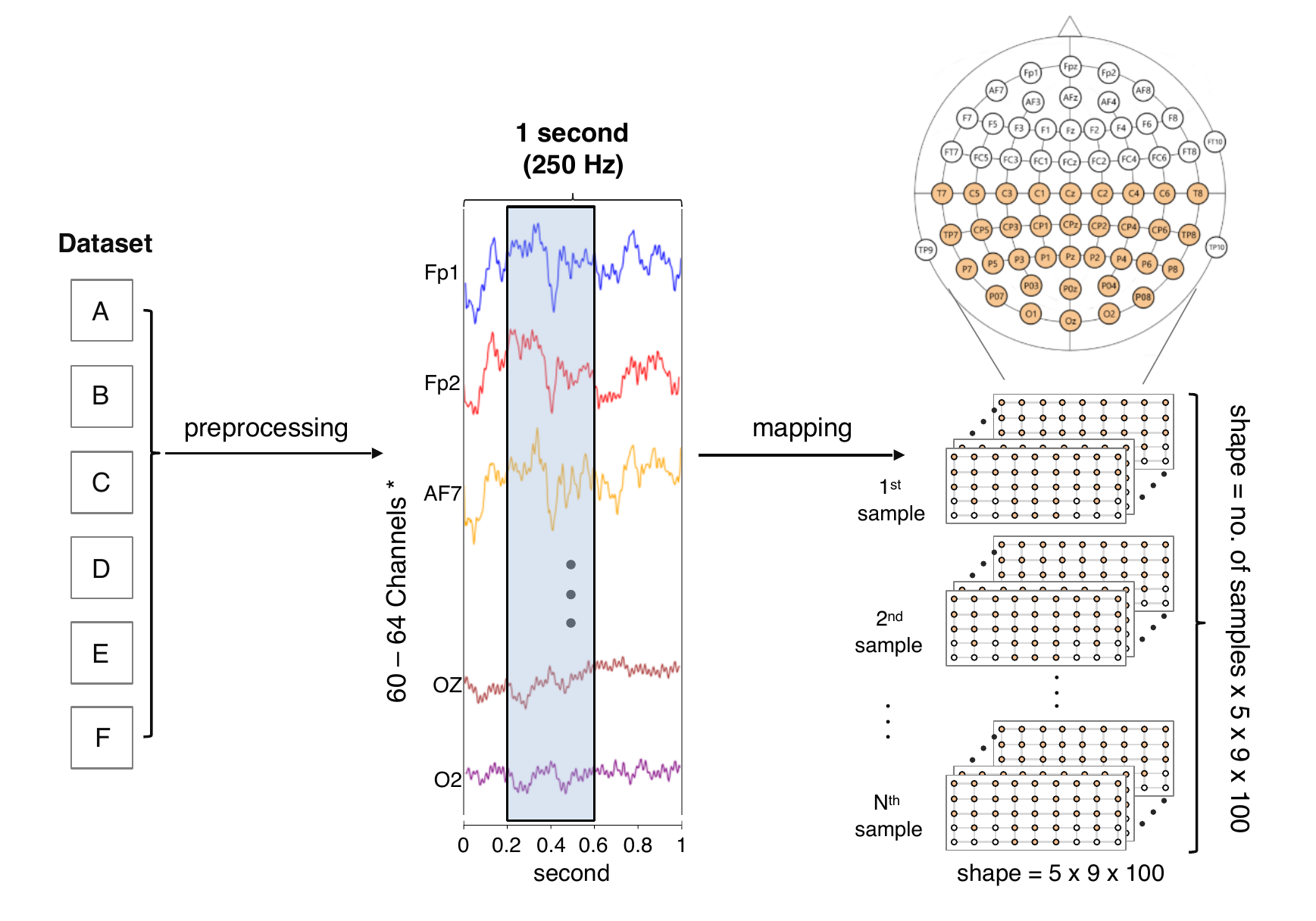}
  \caption{ An overview of our data pre-processing. All data were aggregated from six datasets that differ in number of channels and sampling rates (as shown in Table \ref{datatable}). The data in each dataset was preprocessed (described in Section III) and partitioned into attended and unattended events, labeled in one-second lengths. Only samples measuring from 0.2 to 0.6 seconds were selected, resulting in $100(250Hz\times0.4s)$ points per recording (as described in Section III). Only the midline and occipital parts of the scalp were included (indicated by the orange circles in the figure) and mapped into a 5 x 9 x 100 matrix. Finally, these were used as inputs for the multi-task AE. }
  \label{fig:montage}
\end{figure*}
\begin{figure*}
 \centering
 \includegraphics[width=0.9\linewidth]{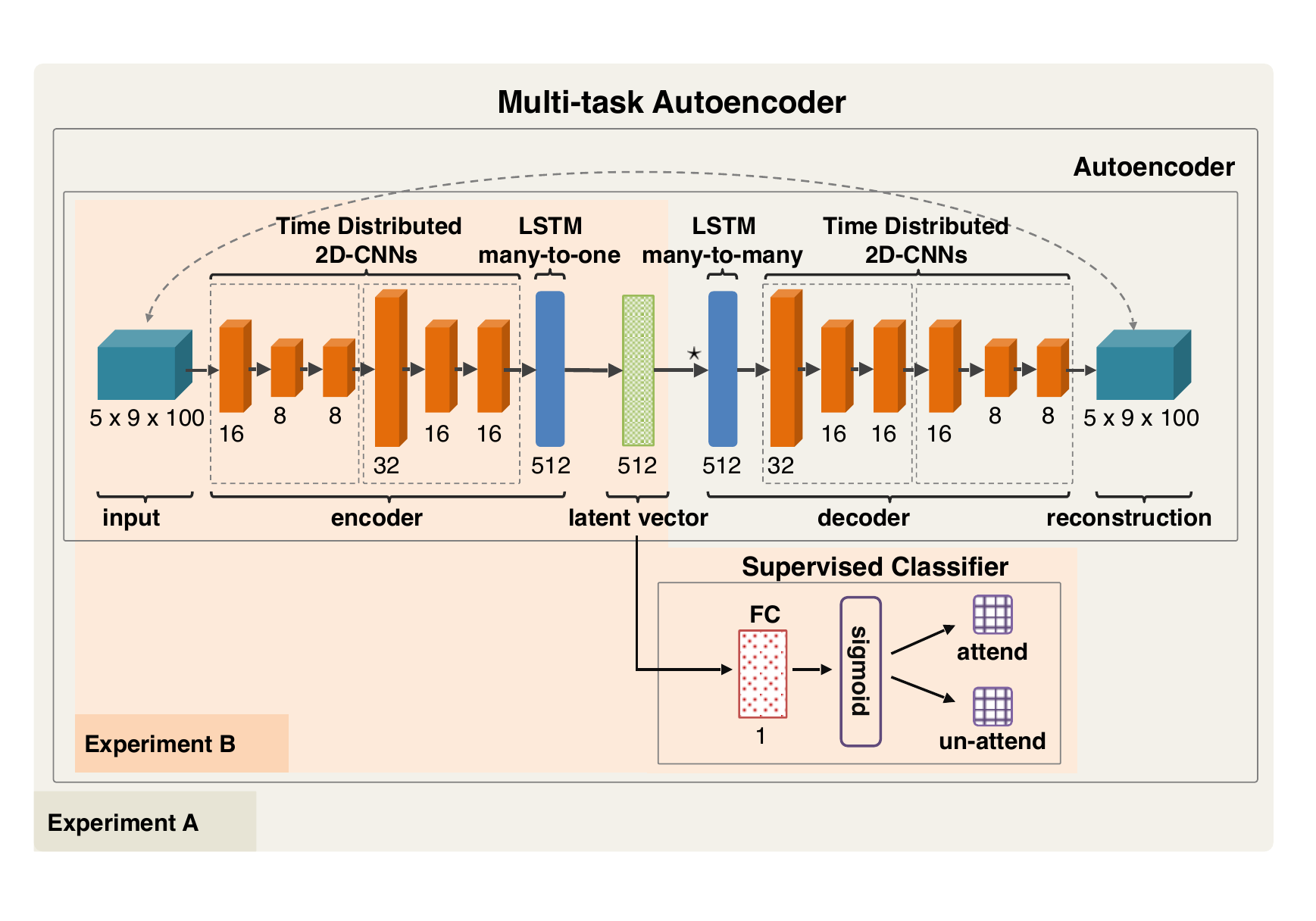}
 \caption{The multi-task AE is composed of two main parts. The first part is the AE containing a encoder network(ERPENet) and a decoder network. The encoder of the AE includes two blocks of CNNs followed by an a LSTM layer in a many-to-one setup. From the LSTM output, a latent vector of the AE was obtained. At $\star$, the latent vector was then repeated 100 times and used as input to the decoder of the AE. The decoder has symmetrical CNN blocks and LSTM sizes, but the number of filters in each layers is different, as shown in the figure. The LSTM in the decoder is also set to return the full sequence (many-to-many). The output of the decoder represents a reconstruction of the input signal. The second part of our model is the supervised classifier. The latent vector from the first part was fed into a single FC layer using sigmoid activation in order to classify the data into two classes: attended and unattended events.}
  \label{fig:AE}
\end{figure*}

\begin{table*}[h]
	\centering
   \caption{The configuration of the multi-task autoencoder architecture}
   \label{modeltable}
   \begin{tabular}{|| c | c | c || c | c | c || c | c | c ||} 
   \multicolumn{3}{c}{Encoder(ERPENet)} & \multicolumn{3}{c}{Decoder}&\multicolumn{3}{c}{Latent Supervised Classifier}\\
  \hline
  Layer & Parameters* & Output & Layer & Parameters** & Output & Layer & Parameters & Output\\
  \hline\hline
  Input & - & (100,5,9,1) & Latent & - & 512 & Latent & - & (512)\\
  \cline{1-3}
  CNN2D & (16,2,2) & (100,3,5,16) & Repeat & 100 & (100,512) & FC & 1 & (1)\\
  BN & - & (100,3,5,16) & LSTM & 96 & (100,96) & Sigmoid & - & 1\\
  \cline{7-9}
  LReLU & 0.10 & (100,3,5,16) & Reshape & - & (100,2,3,16)\\
  \cline{4-6}
  Dropout & 0.20 & (100,3,5,16) & Upsampling & (2,2) & (100,4,6,16)\\
  CNN2D & (8,1,1) & (100,3,5,8) & ZeroPadding & (1,1) & (100,5,7,16)\\
  BN & - & (100,3,5,8) & CNN2D & (32,True) & (100,3,5,32)\\
  LReLU & 0.10& (100,3,5,8) & BN & - & (100,3,5,32)\\
  Dropout & 0.20& (100,3,5,8) & LReLU & 0.10& (100,3,5,32)\\
  CNN2D & (8,1,1)& (100,3,5,8) & Dropout & 0.20& (100,3,5,32) \\
  BN & -  & (100,3,5,8) & CNN2D & (16,False) & (100,3,5,16)\\
  LReLU & 0.10 & (100,3,5,8) & BN & - & (100,3,5,16) \\
  Dropout & 0.20 & (100,3,5,8) & LReLU & 0.10& (100,3,5,16)\\
  \cline{1-3}
  CNN2D & (32,2,2) & (100,2,3,32)& Dropout & 0.20& (100,3,5,16)\\
  BN & - & (100,2,3,32)& CNN2D & (16,False) & (100,3,5,16)\\
  LReLU & 0.10 & (100,2,3,32)& BN & - & (100,3,5,16)\\
  Dropout & 0.20 & (100,2,3,32)& LReLU & 0.10& (100,3,5,16)\\
  CNN2D & (16,1,1) & (100,2,3,16)& Dropout & 0.20& (100,3,5,16)\\
  \cline{4-6}
  BN & - & (100,2,3,16)&Upsampling & (2,2) & (100,6,10,16)\\
  LReLU & 0.10 & (100,2,3,16) & ZeroPadding & (1,1) & (100,7,11,16)\\
  Dropout & 0.20 & (100,2,3,16) & CNN2D & (16,True) & (100,5,9,16)\\
  CNN2D & (16,1,1) & (100,2,3,16) & BN & - & (100,5,9,16)\\
  BN & -  & (100,2,3,16) & LReLU & 0.10& (100,5,9,16)\\
  LReLU & 0.10 & (100,2,3,16) & Dropout & 0.20& (100,5,9,16)\\
  Dropout & 0.20 & (100,2,3,16) & CNN2D & (8,False) & (100,5,9,8)\\
  Flatten & - & (100,96)  & BN & - & (100,5,9,8) \\
  LSTM & (512,0.20) & Latent(512)  & LReLU & 0.10& (100,5,9,8)\\
   \cline{1-3}
  \multicolumn{2}{c}{}  & & Dropout & 0.20& (100,5,9,8)\\
 \multicolumn{2}{c}{}  & & CNN2D & (8,False) & (100,5,9,8)\\
  \multicolumn{2}{c}{}  & & BN & - & (100,5,9,8) \\
  \multicolumn{2}{c}{}  & & LReLU & 0.10& (100,5,9,8)\\
 \multicolumn{2}{c}{}  & & Dropout & 0.20& (100,5,9,8)\\
  \multicolumn{2}{c}{}  & & Reconstruct layer & - & (100,5,9,1)\\
   \cline{4-6}
  \end{tabular}
   \caption*{ *Parameters format in the encoder:  CNN(number of filters,vertical stride,horizontal stride) with filter size of (3,3), LReLU(alpha), Dropout(Dropout Rate), LSTM(filter size,Dropout Rate)\\
   **Parameters format in the decoder: CNN(number of filters,zero padding) with filter size of (3,3)), LReLU(alpha), Dropout(Dropout Rate), LSTM(filter size,Dropout Rate) with full sequence output, Upsampling(vertical scale,horizontal scale), Zeropadding(vertical padding, horizontal padding)}
\end{table*}

\section{Background}
In this section, we first introduce CNN and LSTM, two Deep Neural Network (DNN) layers that are the backbones of ERPENet, in Section A and B. Then, the concept and previous works of AE is described in section C.

\subsection{Convolutional neural network}

CNN is a grid-like topology neural network with a convolution operation at its core. It is proficient at extracting spatial information; such is the basis of many state-of-the-art DNNs in various vision models such as VGG16\cite{simonyan2014very}, Resnet\cite{krizhevsky2012imagenet}, Alexnet\cite{he2016deep}, R-CNN\cite{ren2015faster} because of its parameters sharing and equivalent representations properties. Furthermore, CNN is also adopted in time-series related applications, such as sentence-level classification\cite{kim2014convolutional} that uses CNN on top of word2vec, and similarly in this work. Most CNNs contain an assortment of four different layers: convolutional layer, pooling layer, fully connected layer, and activation layer.

The convolution layer consists of small learnable kernels that are convolved over the entire input space to linearly transform the input, providing translation invariance.

The pooling layer is often placed after the convolution layer to downsample the spatial size of incoming data, thereby reducing the number of computation when stacking multiple convolution layers. 

The Fully Connected (FC) layer is a traditional neural network and usually included at the final part of the CNN, where high-level representations are extracted, to classify the input. Unlike the convolutional layer,  the linear transformation is not restricted by spatial invariance constraints. 

The activation layer is typically a non-linear function which allows the stacking of linear layers.  There are many kinds of functions used for the activation function, however, the one commonly used between convolution layers are the rectified linear unit (ReLU). ReLU is a non-saturating activation function, discarding inputs with negative values by setting them to zero. Another vital activation used in binary classification task is the Sigmoid function. 

\subsection{Long short-term memory}

Long short-term memory (LSTM) is a type of recurrent neural network (RNN), which has the capacity of sequence prediction from self-feedback. Each RNN node has its own internal memory which can produce arbitrary sequences, but RNN suffers from the vanishing and exploding gradient problems. To alleviate these problems, long short-term memory (LSTM) was developed\cite{hochreiter1997long} by adding three gates inside the RNN cell. The three gates are the input gate ($g_{i}^{(t)}$), the forget gate ($f_{i}^{(t)}$) and the output gate ($q_{i}^{(t)}$). Intuitively, the input gate controls the flow of new information entering the cell. The forget gate then decides on which information should be kept in the cell, and the output gate decides when to generate the output. Every gate is based on the state unit($s_i^{(t)}$). Mathematically, it can be represented using the following equations.

\begin{equation} 
f_i^{(t)} = \sigma(b_i^f + \sum_{j}U_{i,j}^fx_j^{(t)} + \sum_{j}W_i^fh_j^{(t-1)})
\end{equation}
\begin{equation}
g_i^{(t)} = \sigma(b_i^g + \sum_{j}U_{i,j}^gx_j^{(t)} + \sum_{j}W_i^gh_j^{(t-1)})
\end{equation}
\begin{equation}
s_i^{(t)} = f_i^{(t)}s_i^{(t-1)} + g_i^{(t)}\sigma(b_i + \sum_{j}U_{i,j}x_j^{(t)} + \sum_{j}W_ih_j^{(t-1)})
\end{equation}
\begin{equation}
h_i^{(t)} = tanh(s_i^{(t)})q_i^{(t)}
\end{equation}
\begin{equation}
q_i^{(t)} = \sigma(b_i^o + \sum_{j}U_{i,j}^ox_j^{(t)} + \sum_{j}W_i^oh_j^{(t-1)})
\end{equation}
 
where  $U$ is the weight matrix connecting the inputs to the current hidden layer. $W$ is the weight matrix connecting the previous hidden layer and current hidden layer. $b$ is the bias matrix. $x(t)$ is the current input vector and $h_i^{(t)}$ is the current hidden layer vector, where $i$ denotes a dependent cell, and $t$ denotes the time step in each cell. $\sigma$ is the sigmoid function acting as a gate in the LSTM unit.

\subsection{Autoencoders}

Autoencoders (AEs) were first introduced in the 1980s for unsupervised feature extraction\cite{rumelhart1985learning}, dimensionality reduction\cite{hinton2006reducing} and data compression\cite{theis2017lossy}. AEs have two main components: the encoder and the decoder. The encoder network $(E)$ maps an input signal ($s$) to a latent representation ($h$) and the decoder network $(D)$ tries to reconstruct the input signal at the output layer using the latent representation. Ideally, there should be fewer nodes in the latent representation than in the input, resulting in the creation of a bottleneck effect, limiting the information passed to the decoder network. The network is trained to minimize reconstruction loss, defined as $L(s,D(E(s))$. In a complex supervised learning models\cite{vincent2010stacked}, AEs was used to initialize the weights one layer at a time by minimizing the reconstruction loss.

In a recent work \cite{FengScarce}, AE was previously adopted to pre-train the classifier network for EEG. They proposed deep AE for solving the problems on EEG datasets with scarce label information. The AE was first pre-trained using unlabeled data to extract EEG features in an unsupervised manner. Then, the encoder network was attached with a classifier and further trained with labeled data. Another work \cite{IoBT} created an optimized framework for seizure prediction using huge EEG datasets. In their framework, AEs played an important role as an effective data compression method. However, most studies only applied typical stacked-AEs or modified-AEs (denoising sparse AE) through unsupervised learning to extract the features. The learned features were then passed through various classification methods for the target task as presented in \cite{access1,denoise,hmm,access2}. As stated in the introduction, we proposed the a multi-task AE, consisting of both 2D-CNNs and LSTMs, to perform feature extraction, data compression and classification on large-scale EEG datasets, which are discussed in the following sections.

\section{Methodology}

In section A, we first provide the motivations of incorporating CNN and LSTM into the AE. Then, the architecture of the multi-task AE is described in Section B, followed by its loss function in Section C.

\subsection{Integrating CNN and LSTM into AEs}
In the recent years, modern AEs have become more complex due to the integration of various kinds of layers into the encoder network and decoder network. In this work, the encoder and decoder are comprised of CNNs to learn the high-level features in spatial domain of multi-channel EEG and LSTM to learn the temporal relationship.

Our proposed model is trained on a combination of six public datasets, which have different electrodes poisonings, different resolution and different sampling rate. This creates a need for a network architecture that can handle these differences. The EEGs are recorded by using the scalp electrodes placement known as 10-20 system standard\cite{Soutar201419}. In higher resolution recordings of EEG, a Modified Combinatorial Nomenclature (MCN) was developed by adding 10\% divisions to increase the EEG channels creating discrepancy across multiple resolutions. To overcome the inconsistency in electrode positioning across multiple datasets, 2D-CNNs are included at the beginning of the network to extract the spatial information in our proposed model.

Each EEG channel is mapped directly on 2D-grid topography representing the 2D scalp-map to preserve spatial features, as shown in mapping procedure in Figure \ref{fig:AE}. Since we are focusing on P300 tasks, we select only the EEG channels between midline and occipital (marked with orange color in Figure \ref{fig:montage}), totaling 35 channels(Cz, C1-6, T7,8, TP7,8, CPz, CP1-6, Pz, P1-8, POz, PO3,4,7,8, O1,2 and Oz). These were previously reported as the optimal set of channels in BCI P300\cite{rakotomamonjy2008bci}. 

However, CNN is known to lack the ability to capture the long-term relationships within time series, since only the data points within the CNN filters are used for information extraction. To combat this problem, the LSTM is incorporated into our proposed model. A LSTM layer is included after the 2D-CNNs layers, to learn the long-term representation of the sequential data and compress data in the temporal dimension. The mechanism of LSTM provides an efficient method to encode the representation of EEG into a single vector, which can be decompressed in the decoder network or be used as input to a related task. 

\subsection{Multi-task autoencoder} 

The multi-task autoencoder model can be split into three networks: encoder network, decoder network and supervised classifier network as shown in Figure \ref{fig:AE} and more details of each layer in the Table \ref{modeltable}. However, they are trained simultaneously to reconstruct the input in the decoder network and to classify the input in the supervised classifier network.

The encoder network is composed of 2 CNN blocks followed by a LSTM layer. The arrangement of CNN blocks and its composition, including three time-distributed CNNs, a Batch Normalization(BN) layer, a Leaky Rectified Linear Unit(LReLU)\cite{maas2013rectifier} and a dropout layer , was inspired by the VGG16 architecture\cite{simonyan2014very}. The final CNN block is connected to a many-to-one LSTM. The input has a dimension of (100,5,9,1), formatted by time, vertical coordinate, Horizontal coordinate, data.

Each CNN block has three stacked CNNs, with a larger number of filters and a stride of 2 in the first CNNs. The filter sizes in the network were optimized by gradually decreasing from a larger number until a degradation in performance was observed. The layers parameters are listed in the Table \ref{modeltable}. A stride of 2 is applied to reduce the size of the model, in a similar way to max pooling. Max pooling was avoided in this study because there was a report of checkerboard artifact that generated high-frequency noise in the reconstruction output\cite{odena2016deconvolution}.  The latter two CNNs extract the information without any additional compression. After every CNN layers, dropout regularization with a 0.20 dropout rate was applied prior to feeding into the next CNN layer to avoid overfitting. The dropout rate was recommended to be increased if overfitting was observed, but it should be between 0.2 and 0.5. In this work, the dropout rate of 0.2 was used throughout the network since there was no severe overfitting observed. An output of the last CNN block has a shape of (100,96), which is fed into a LSTM layer, comprising 512 LSTM units with a recurrent dropout. The output of the LSTM at the final time step is considered as the latent vector, encapsulating the compressed information. Because the size of the LSTM unit is also the size of latent vector, which we want to minimize while maintaining data representation, 512 is a number that we found robust towards multiple P300 tasks. As a result, the EEG signal is compressed in the spatial and temporal domains into a single vector of size 512.

In the decoder network, all layers are aligned symmetrically with the encoder network. The latent vector is repeated 100 times to construct the data in a temporal format, matching the input format required for the many-to-many LSTM. In the CNN blocks of the decoder network, upsampling and zero-padding layers are added to reconstruct the network. Zero-padding layer and zero-padding option in CNNs are used to reverse the dimensional reduction by convolution kernel without padding in the encoder network. Similarly to the encoder section, we decide not to use deconvolution, because of the reported checkerboard artifact generated\cite{odena2016deconvolution}. After two blocks of CNN, the input EEG signal is reconstructed.

To prevent the AE from over-compressing the EEG, a supervised portion was added - extended from the latent vector. Thus, the model is capable of learning to classify the input signal along with that of the reconstruction. A basic supervised classifier was added, composed only of a single FC unit and sigmoid activation after the latent vector, which could be considered as an auxiliary input to the model. This extended supervised network creates a constraint for the latent vector to become interpretable as well as a compression of EEG. The encoder network is a transferable model called ERP Encoder Network (ERPENet).  Here, we proposed a multi-task AE with CNN and LSTM model to reconstruct and classify EEG signals (Figure \ref{fig:AE}).

\subsection{Loss Function}
Our proposed multi-task AE model was trained for two different tasks: reconstruction and binary classification. The Two loss functions, which were implemented in TensorFlow\cite{tensorflow2015-whitepaper}, were incorporated and combined with a weight ($\beta$) on the classification loss function.  

For the reconstruction loss, the Mean Square Error (MSE) metric was computed from the difference between the reconstruction and the input. Due to the fact that most channel mapping is blank, the MSE function needs to be modified to compute the reconstruction loss only on the feature ($x_j$) not filled with zero as in Eq. 6.
\begin{equation}
L_{MSE}(s) = \|s_j-D(E(s))_j\|^2, s_{j}\neq0
\end{equation}
$s_j$ is the input signal where j denotes the channel containing blanks on 2D mapping. This prevents the reconstruction preferring to output zeros in the early training stage. 

In the latent supervised classifier, attended and unattended events are classified by the sigmoid binary cross-entropy as shown in Eq. 7. 
\begin{equation}
L_{Binary}(y,y')=-\frac{1}{n}\sum_{i=1}^{n}y_i \ln (y'_i)
\end{equation}
where $n$ denotes the total number of the input signals$(s)$. The prediction ($y'$) was predicted straight from the sigmoid attached to the $f(x)$ while $y$ is a true label. Binary class weights ($W_{c,i}$), where $c$ denotes class of sample $i$, was optimized to penalize the imbalances classes. $\beta$ is introduced to weight the classification loss to the reconstruction loss. To find $\beta$, grid search over a set, $[0.25,0.333,0.5,0.667,0.75]$, has been performed to minimize $L_{total}(s,y,y')$ which $beta$ equals to 0.667 yielded an optimal result. Finally, the total loss function was a summation of both objective loss functions weighted by dataset weights $(W_{d,i})$,where $d$ denotes the sample of dataset $i$.

\begin{equation}
L_{total}(s,y,y') = W_{(d,i)} [\beta W_{(c,i)}L_{Binary}(y,y') + L_{MSE}(s)]
\end{equation}

\begin{table*}[h]
\centering
\caption{Datasets used to train ERPENet}
\label{datatable}
\resizebox{\linewidth}{!}{%
\begin{tabular}[t]{lccccclcc}
\hline
\multicolumn{1}{c}{\textbf{dataset}} & \textbf{stimuli type} & \textbf{no. of subjects} & \textbf{no. of samples} & \textbf{sampling rate (Hz)} & \textbf{no. of channels} & \multicolumn{1}{c}{\textbf{channels}} & \textbf{EEG recording system} & \textbf{bandpass filter} \\ \hline
\makecell[l]{ Exploiting P300 amplitude changes \\ (P300-Amplitude) } & visual & 12 & 26182 & 2048 & 64 & \makecell[l]{ Fpz, Fp1,2, AFz, AF3,4 \\ AF7,8, Fz, F1-8, FCz, \\ FC1-6, FT7,8, Cz, C1-6, \\ T7,8, TP7,8, CPz, CP1-6, \\ Pz, P1-10, POz, PO3,4,7,8, \\ Oz, O1,2, Iz } & \makecell[c]{ BioSemi  ActiveTwo } & 0.15--5 Hz \\ \\
\makecell[l]{ BCI Competition III - Dataset II \\ (BCI-COMP) } & visual & 2 & 8295 & 240 & 64 & \makecell[l]{Fpz, Fp1,2, AFz, AF3,4,7,8, \\ Fz, F1-8, FCz, FC1-6, \\ FT7,8, Cz, C1-6, T7,8,10, \\ TP7,8, CPz, CP1-6, Pz, \\ P1-8, POz, PO3,4,7,8, \\ Oz, O1,2, Iz } & not specified & 0.1--60 Hz \\ \\
\makecell[l]{ Auditory multi-class BCI \\ (Auditory-BCI) } & auditory & 10 & 40161 & 250 & 60 & \makecell[l]{ Fp1, MasL,R, AF3,4, Fz, F1-8, \\ FCz, FC1-6, FT7,8, Cz, \\ C1-6, T7,8, CPz, CP1-6, \\ TP7,8, Pz, P1-10, POz, \\ PO3,4,7,8, Oz, O1,2 } & \makecell[c]{Ag/AgCl electrodes, \\ amplifier from  Brain  Products} & 0.1--250 Hz\\ \\
\makecell[l]{ BCI-spelling using RSVP \\ (BCI-Spell) } & visual & 13 & 84360 & 250 & 63 & \makecell[l]{ Fp2, AF3,4, Fz, F1-10, FCz, \\ FC1-6, Cz, C1-6, T7,8, \\ CPz, CP1-6, TP7,8,Pz, \\ P1-10, POz, PO3,4,7-10, \\ Oz, O1,2, Iz, I1,2 } & \makecell[c]{actiCAP active electrode system \\ from Brain Products} & 0.016--250 Hz\\ \\
\makecell[l]{ Examining EEG-Alcoholism Correlation \\ - control group \\ (EEG-Alcohol) } & visual & 122 & 10962 & 256 & 61 & \makecell[l]{ Fpz, Fp1,2, Fz, F1-8, FT7,8, \\ AFz, AF1,2,7,8, FCz, \\ FC1-6, T7,8, Cz, C1-6, CPz, \\ CP1-6, TP7,8, Pz, P1-8, \\ POz, PO1,2,7,8, Oz, O1,2 } & not specified & 0.02--50 Hz \\ \\
\makecell[l]{ Decoding auditory attention \\ (Decode-Audi)} & auditory & 11 & 24560 & 200 & 63 & \makecell[l]{ Fp1,2, AF3,4,7,8, Fz, F1-10, \\ FCz, FC1-6, FT7,8, \\ Cz, C1-6, T7,8, TP7,8, \\ CPz, CP1-6, Pz, P1-10, \\ POz, PO3,4,7,8, Oz, O1,2 } & \makecell[c]{actiCAP active electrode system \\ from Brain Products} & 0.016--250 Hz\\ \hline
\end{tabular}
}
\end{table*}

\section{Datasets and data preprocessing}

\subsection{Datasets}

	The following six datasets were incorporated in this study. All were from the P300-BCI experimental tasks based on the oddball paradigm, including visual and auditory stimuli, each of which has their own specific attended and unattended events. Moreover, these datasets were chosen in order to represent the strength of the model in which the neural networks, especially the CNN, are still able to capture the essential features from the input signals even though they were derived from various montage systems, sampled at different sampling rates, collected with diverse hardware filtering methods, and recorded in an unequal number of channels. Descriptions of the datasets are shown in Table \ref{datatable} as follows:
\begin{itemize}
  \item Exploiting P300 Amplitude changes \cite{citi2010}: this dataset resulted from a visual stimuli experiment. The study aimed at identifying the factors limiting the performance of BCIs based on ERPs, in order to improve the transfer rate and usability of these interfaces. In every run, each subject was asked to look at a 6x6 matrix, including 36 different characters. The rows and columns of the matrix were randomly highlighted one at a time for a short period, specifying a target character before each run. Each subject was then asked to mentally count the number of times any row or column, including the target character, intensified. During the experiment, EEG signals were collected using a BioSemi ActiveTwo EEG system. Subsequently, the signals were bandpass filtered in the band 0.15-5 Hz.	
  \item BCI Competition III - Dataset II \cite{rakotomamonjy2008bci}: this study used a visual stimulus with an intra-subject classifier proposed to predict the desired character from EEG signals. The experiments were similar to those in the above dataset. They also used the 6x6 matrix and attended events also occurred when any row or column with the target character was flashed. Instead of focusing on one character for each run, participants in this experiment were asked to focus on a single word containing a sequence of five characters. For each character epoch, rows and columns were randomly highlighted 180 times (6 rows x 15 times and 6 columns x 15 times), 30 of which included the target character specified as an attended event. For every run, each subject had to monitor five characters per epoch. Signals from the subjects were collected using a montage system not specified in the paper. Finally, they were bandpass filtered from 0.1-60 Hz.
  \item auditory multi-class BCI \cite{schreuder2010}: ]: this dataset was collected from a study using auditory experiments. They tried to propose a multi-class auditory-BCI classification using spatially distributed, auditory cues. In the experiment, each participant was surrounded by eight speakers, only five of which were used. These speakers were programmed to turn on at random, one at a time. In each run, a subject was instructed to mentally keep track of the extent to which the target direction (target speaker) was stimulated. The EEG was recorded using a number of Ag/AgCl electrodes, amplified using a 128-channel amplifier from Brain Products, filtered by an analogue bandpass filter between 0.1 and 250 Hz.
  \item BCI-Spelling using Rapid Serial Visual Presentation (RSVP) \cite{acqualagna2013}: the aim of this study was to develop a visual speller that did not require eye movements to overcome the limitations of conventional BCIs. Each subject participated in two experiments: In the first experiment, geometric shapes were randomly flashed on the screen. Each geometric contained a unique set of five characters and had a unique shape and color. In the second experiment, each shape was changed to contain only a single character. For every run in both experiments, each participant was asked to mentally count the number of times the target character was shown on the screen. The signals were recorded using an actiCAP active electrode system from Brain Products (Munich, Germany). All skin electrode impedances were kept below 20 k\si{\ohm}. The bandpass of the hardware filter was 0.016-250 Hz.  
  \item Examining EEG-Alcoholism Correlation  \cite{UCI}: This data was obtained from a large study of visual stimuli experiments. There were two groups of subjects: alcoholic and control (healthy subjects). Only control subjects were selected in order to avoid outlier samples which may be affected by alcoholism. Each subject participated in two experimental sets. Firstly, a single stimulus (S1) of one picture was randomly shown on the screen. The attended events were when the target picture was shown. Secondly, two pictorial stimuli (S1 and S2) were shown at the same time and the attended events were when both were identical. During the experiments, each subject was attached with a 61-lead electrode cap (Electro-Cap International). The impedances were kept below 5 k\si{\ohm}. The signals were filtered with a bandpass of 0.02-50 Hz, and recorded on a computer with subsequent 32 Hz low pass digital filtering.
  \item Decoding auditory attention \cite{bnci15}]: This experimental study of auditory stimuli attempted to prove the concept that paying attention to a particular instrument in polyphonic music (music with several instruments playing in parallel) can be inferred from EEG. In the experiment, each subject listened to four different types of polyphonic music clips in which each clip included three types of instruments. Before each clip, one out of three instruments were specified as a target instrument. Normally, all instruments played simultaneously in a repetitive simple pattern. Each subject was asked to attend to the target instrument and count the number of times the pattern deviated from the original. The EEG signals were recorded using an actiCAP active electrode system from Brain Products (Munich, Germany). All skin electrode impedances were kept below 20 k\si{\ohm}. The bandpass of the hardware filter was 0.016$-$250 Hz.
\end{itemize}

\subsection{Data preprocessing}
Before using these datasets to evaluate our method, power line noise (50 Hz) was manually checked, and a Notch filter was applied if a noise was found. A low pass second order Butterworth filter of 30 Hz and high pass second order Butterworth filter of 0.5 Hz were then applied to normalize the datasets. We used a low order filter because some of the datasets had already been preprocessed. For consistency, they were resampled from the original sampling rates to 250 Hz using Fourier method.

From previous works\cite{lin2018novel}\cite{kubler2009brain}, the P300 interest period was shown to be between 0.2 and 0.6 seconds after the stimulus. Therefore, we reduced the length of the EEG signal to 0.4 seconds.

\section{Experimental evaluation}
In this section, the evaluation methods were constructed to examine the properties of proposed multi-task AE model in two different perspective. The first evaluation (Experiment A) measured the performance of trained multi-task AE model on an unseen dataset, by comparing reconstruction error and classification accuracy with a previous work. In Experiment B, weights in encoder network was taken from Experiment A as a pre-trained network and continued the training on an unseen P300 dataset to classify attended, and unattended events.    

\subsection{Experiment A: multi-task AE reconstruction error}
In the first experiment, the compression performance of the proposed multi-task AE was measured by reconstruction error and attended/unattended events classification accuracy on an unseen dataset. This evaluation method was designed to test the robustness of our proposed model over multiple P300 datasets. One dataset was held back from the multi-task AE training and used as a testing dataset to evaluate the trained AE. Excluding the testing dataset, five datasets were aggregated, stratified, and randomly split into two sets: training and validation, with the ratio of 90:10. Here, one dataset was held back from the training, as a testing set, to demonstrate the robustness of our proposed multi-task AE model across multiple unseen datasets.

Our input data contains 0.4 seconds of 35 EEG channels (Cz, C1-6, T7,8, TP7,8, CPz, CP1-6, Pz, P1-8, POz, PO3,4,7,8, O1,2 and Oz). The data were compressed via the encoder network into a latent vector of size 512. Hence, the compression ratio of our proposed AE model was $6.84 (0.4s \times 250 Hz\times35 /512)$. 

To compare the model capabilities for compression and classification, we tested our model against the multi-task Stacked Label Consistent Autoencoder (SSLC-AE)\cite{gogna2017}, proposed by Hoffmann et al. The baseline shared a certain level of similarity with our proposed model. Specifically, the SSLC-AE contained AE and a supervised network and was trained as a multi-task model. However, the reconstruction and classification losses in the SSLC-AE model were not combined into a single loss but kept separate and applied alternatively with the Split Bregman technique, instead of the reverse-mode auto differentiation used in our model. Unlike our proposed model, the SSLC-AE model contains only two FC layers in the encoder and decoder networks. FC is known to be sensitive to the training data and may easily overfit, limiting the size of SSLC-AE model to be small and shallow. To overcome this problem, they augmented the training data to avoid the overfitting problem while our proposed model was constructed by CNNs and LSTM, which tends to make the model robust to overfitting than FC as described in Section II, and trained using aggregated EEG from multiple datasets, which is an alternative way to combat the limited data problem.

For the original SSLC-AE, the encoder network is composed of 2 FC layers, 125 nodes and 63 nodes, with a symmetrical decoder and supervised classification on the latent vector. However, for a fair comparison with our proposed model, we increased the number of nodes in FC layers to 500 and 250, respectively. As such, the latent vector in SSLC-AE was comparable to that in our proposed AE model. The 2D-grid map was flattened to a single vector and used as the input of SSLC-AE. The reconstruction error of SSLC-AE was also modified to compute only on the non-blank inputs. Prior to the training of SSLC-AE, optimization in the SSLC-AE model was tuned with the same grid search as our proposed model. The RMSprop optimizer with a $2^{-6}$ learning rate and a decay rate of $10^{-5}$ yielded the lowest loss in all dataset permutations. The AE was trained until there was no improvement in validation loss for 100 epochs.

In the multi-task AE training, the optimization algorithm and learning rate were chosen by grid searching between RMSprop\cite{Tieleman2012} and vanilla SGD\cite{kiefer1952stochastic} , with the initial learning rate between $[2^{-10}, 2^{-5}]$ in the decreasing power of 2, and the decay rate between $[10^{-7}, 10^{-4}]$ in the decreasing power of 10. After grid searching all six permutations, vanilla SGD with a learning rate of $2^{-8}(0.002)$ and a decay rate of $10^{-5}$ achieved the lowest validation loss. The AE was trained until there was no improvement in the validation loss for 100 epochs.

Another objective of our proposed model was to use the latent vector to predict ERP attended and unattended events directly without any additional neural network layers besides the sigmoid. For each test dataset, we train a supervised classifier using 10-fold cross-validation. Eight folds were used in training. One was used as a validation set to tune the network. The last fold was used to test the algorithm. The supervised classification network (a single FC node) was trained using the features extracted from the encoder. In the results, only attended and unattended event classes were predicted from the supervised network.

\subsection{Experiment B: Adoption of pre-trained model(ERPENet)}
Instead of using the encoder as a fixed feature extractor, in this experiment we trained (fine-tuned) the pre-trained ERPENet on an unseen P300 dataset to only classify attended and unattended events as a supervised learning task. The primary objective of the pre-trained network was to partially combat a drawback of DL that requires a substantial amount of data in the training process, as an alternative approaching to data augmentation mentioned in Experiment A.

For a small single dataset, DL performance was often outperformed by the traditional machine learning algorithms. Our proposed multi-task AE was trained with a variety of large EEG datasets, which partially solved the problem.  

General representations of EEG was learned via multi-task training, thereby minimizing the overfitting problem. In this section, the trained weights of CNN blocks and LSTM in the encoder network have been adapted and extended using a supervised classifier as shown in Figure \ref{fig:AE}.

\begin{figure}[H]
  \centering
  \captionsetup{justification=centering}
  \scalebox{0.6}{\input{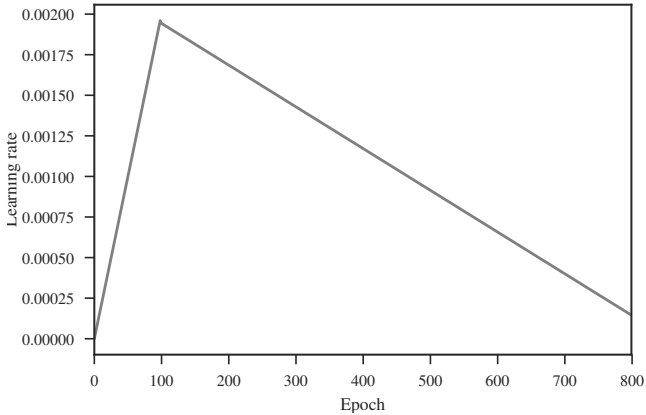}}
  \caption{The learning rate used in training of the pre-trained model. Learning rate gradually increases from 0.00002 to 0.002 at epoch 100, and linearly decreases until reaching 0.0002 at epoch 800}
  \label{fig:learningRate}
  
\end{figure}

First, we compared the training losses of an adapted pre-trained model and the same model with Xavier\cite{glorot2010understanding} initialized weights (random initialization). In the adapted model, a concave-down triangular learning rate technique was applied as shown in Figure \ref{fig:learningRate}\cite{smith2017cyclical}. The model with Xavier initialized weights was trained using the SGD with learning rate of $2^{-8}(0.002)$ and the decay rate of $10^{-5}$ as in the Experiment A. 

To further validate our model using a state-of-the-art P300 feature extraction method, we compared our ERPENet against one traditional P300 feature extraction, the Xdawn algorithm \cite{rivet2009xdawn}, and other three deep learning models: EEGNet\cite{Waytowich2018}, DeepConvNet\cite{hbm23730} and SSLC-AE. Xdawn is one of state-of-the-art\cite{lotte2018review} dimensional reduction algorithms, enhancing the features of ERP-based EEG. Bayesian LDA classification was applied to classify the attended and unattended event classes \cite{mika1999fisher}, with 10-fold cross-validation used to cross-validate the Xdawn-LDA. Eight folds were used in training — one as a validation set to tune the number of components in Xdawn. The last fold was used to test the algorithm as reported in the next section. 

EEGNet\cite{Waytowich2018} model comprises three compact CNNs to classify EEG-based BCIs in various tasks, including P300. Each CNN layer is convoluted in different input dimension to extract the representative features. In this evaluation, we configured the filter sizes to match the model with the best performance reported in \cite{Waytowich2018}, which were eigth temporal filters and two spatial filters per temporal filter. Unlike ERPENet, EEGNet avoids overfitting problems by limiting the trainable parameters instead of dropout regularization, which reduces classification performance in advance tasks. In \cite{Waytowich2018}, the performances of EEGNet in P300 task outperformed Xdawn algorithm and were comparable with DeepConvNet.

DeepConvNet\cite{hbm23730} model comprises five CNNs which is a deeper model than EEGNet. Hence, Dropout layers with a rate of 0.5 are added after CNN layer to prevent the overfitting in the same way as in ERPENet.

SSLC-AE, EEGNet and DeepConvNet models were also pre-trained and fine-tuned with the same technique and hyperparameter optimization as in ERPENet. The parameters in EEGNet and DeepConvNet were configured as recommended in \cite{Waytowich2018} to match inputs in this work.

All models were trained and tested on a machine with NVIDIA P100 GPU, E5-2667 CPU and 128 GB of memory with a batch size of 32 in the training and 128 in the testing. We report the average training and testing times for each trial per epoch over all datasets in second unit in Table \ref{time}. Please note that the total training time($T_total$) in all models depends on the size of dataset, which is $T_{total} = N_{pre} D_{pre}  T_{pre} + N_{fine} D_{fine}  T_{fine}$, where $N_{pre}$ denotes number of training epochs in the pre-training, $N_{fine}$ denotes number of training epochs in the fine-tuning, $D_{train}$ denotes size of pre-trained dataset(5 datasets), $D_{test}$ denotes size of fine-tuning  dataset, which is $80\%$ of holding out dataset as the testing dataset, $T_{pre}$ denoted time used in pre-training and $T_{fine}$ denoted time used in fine-tuning. The number of training epochs depends on the data, which we observed the value to be between 500-800 epochs in the pre-training process and less than 300 in the fine-tuning process. In the fine-tuning process, only the encoder network and classifier network were trained, resulting in the difference between times used in pre-training and fine-tuning.

\begin{table}[H]
\captionsetup{justification=centering}
\caption{Time complexity (seconds per trial per epoch) for all models and number of trainable parameters for all deep learning models}
\label{time}
\resizebox{\columnwidth}{!}{
\begin{tabular}{lcccc}
\hline
\textbf{}
\textbf{Models}& 
\textbf{Pre-training time}& 
\textbf{Fine-tuning time}& 
\textbf{Inference time} &
\textbf{Trainable parameters}\\ \hline
ERPENet & 10.02e-4 & 6.23e-4 & 3.82e-4 & 4,207,871\\
SSLC-AE & 7.52e-4 & 5.73e-4 & 3.65e-4 & 3,756,251\\
EEGNet & 2.91e-5 & 2.91e-5 & 1.55e-5 & 1793\\
DeepConvNet & 3.53e-4 & 3.53e-4 & 1.73e-4 & 154,801\\
XdawnLDA & \multicolumn{2}{c}{-----1.18e-5-----} & 4.39e-6 & - 
\\ \hline
\end{tabular}
}
\end{table}


\section{Results}
In this section, the results of Experiments A and B are shown and statistically analyzed to evaluate our proposed multi-task AE model.

\subsection{Experiment A: multi-task AE reconstruction error}
Our proposed multi-task AE was trained on six datasets using permutation testing and evaluated by reconstruction error(MSE), and classification errors (accuracy and area under the curve). 

In table \ref{resulttable1}, the mean of reconstruction error and its standard error was reported for both the proposed multi-task AE and SSLC-AE. The results relating to the significantly outperformed methods are in bold text. With the Wilcoxon signed-rank test, a two-sided p-value for the null hypothesis where the mean difference of zero is less than 0.01 indicates that the MSEs of the proposed multi-task AE are competitive against SSLC-AE.

\begin{table}[H]
\captionsetup{justification=centering}
\caption{Reconstruction error of the proposed multi-task AE and SSLC-AE on six different datasets trained by holding out the testing dataset}
\label{resulttable1}
\resizebox{\columnwidth}{!}{
\begin{tabular}{lccc}
\hline
\textbf{}                            & \multicolumn{1}{c}{\textbf{Proposed multi-task AE}}                      & \multicolumn{1}{c}{\textbf{SSLC-AE}}                               & \\ \hline
\multicolumn{1}{c}{\textbf{dataset}} & \multicolumn{1}{c}{\textbf{MSE}} & \multicolumn{1}{c}{\textbf{MSE}}  \\ \hline
P300-Amplitude    &   \boldmath$0.1447\pm0.0326$       &      $0.3224 \pm 0.0142$                           \\
BCI-COMP     & \boldmath$0.1325\pm 0.0184$                   & 0.3821 $\pm$  0.0117                \\
Auditory-BCI             &   \boldmath$0.2572\pm 0.0226$              & 0.4206 $\pm$   0.0193                \\
BCI-Spell              & \boldmath$0.2622\pm 0.0214$                 &      0.4399  $\pm$ 0.0184     \\
EEG-Alcochol           &  \boldmath$0.1494\pm 0.0162$           &0.3436 $\pm$   0.0102           \\
Decode-Audi          & \boldmath$0.2761\pm 0.0196$                   & 0.4635 $\pm$   0.0095              \\ \hline
\end{tabular}
}
\end{table}

In the classification evaluation, the area under the curve (AUC) of the receiver operating characteristic (ROC) is reported in addition to classification accuracy (ACC), to test the discriminability of the models. AUC is also known to be insensitive to imbalance classes, validating the binary classification model better than the accuracy metric. 

\begin{table}[H]
\captionsetup{justification=centering}
\caption{Attended and unattended event classification ACC and AUC of the ERPENet and SSLC-AE on six different datasets trained by holding out the testing dataset}
\label{resulttable2}
\resizebox{\columnwidth}{!}{
\begin{tabular}{lllllll}
\hline
\textbf{}                            & \multicolumn{2}{c}{\textbf{ERPENet}}                      & \multicolumn{2}{c}{\textbf{SSLC-AE}}                               & \\ \hline
\multicolumn{1}{c}{\textbf{dataset}} & \multicolumn{1}{c}{\textbf{ACC}} & \multicolumn{1}{c}{\textbf{AUC}} & \multicolumn{1}{c}{\textbf{ACC}} & \multicolumn{1}{c}{\textbf{AUC}}  \\ \hline
P300-Amplitude   & \boldmath $86.32 \pm 2.73$     & \boldmath$80.60 \pm 0.43$   & 81.20 $\pm$ 0.87 & 73.72 $\pm$ 1.53                                  \\
BCI-COMP     & 83.54 $\pm$ 1.53       & 69.15 $\pm$ 0.83       & \boldmath$86.62 \pm 1.4$ & \boldmath$70.97 \pm 2.28$             \\
Auditory-BCI     & $72.56\pm 1.02$      & $54.28\pm2.53$  & 72.43 $\pm$ 1.31      & 50.11 $\pm$ 0.08                  \\
BCI-Spell   & 73.28 $\pm$ 0.58    & 51.55 $\pm$ 1.42       & $76.37\pm1.88$  & $58.47\pm 0.63$\\
EEG-Alcochol    & 63.02 $\pm$ 2.87         & 67.40 $\pm$ 2.83 & \boldmath$73.33 \pm 3.47$  & \boldmath $74.32 \pm 1.08$              \\
Decode-Audi    & 54.76 $\pm$ 1.62      & 50.03 $\pm$ 0.07  & $56.14 \pm 1.45$  & 
$53.82\pm 0.12$ \\ \hline
\end{tabular}
}
\end{table}

\begin{table*}[h]
\centering
\captionsetup{justification=centering,margin=2cm}
\caption{Attended and unattended event classification ACC and AUC of fine-tuned ERPENet, EEGNet, DeepConvNet and SSLC-AE, in comparison with XdawnLDA on six different datasets}
\label{resulttable3}
\begin{tabular}{lllllll}
\hline
\textbf{}                                       & \multicolumn{2}{c}{\textbf{ERPENet}}&
\multicolumn{2}{c}{\textbf{EEGNet}} & 
\multicolumn{2}{c}{\textbf{DeepConvNet}}
\\ \hline
\multicolumn{1}{c}{\textbf{dataset}} & \multicolumn{1}{c}{\textbf{ACC}} & \multicolumn{1}{c}{\textbf{AUC}} & 
\multicolumn{1}{c}{\textbf{ACC}} & \multicolumn{1}{c}{\textbf{AUC}} &
\multicolumn{1}{c}{\textbf{ACC}} & \multicolumn{1}{c}{\textbf{AUC}}\\ \hline
P300-Amplitude        & \boldmath$88.52 \pm 2.42$                   & \boldmath $84.23 \pm 0.98$  & 83.33 $\pm$ 1.30 & 79.24 $\pm$ 2.89    & 83.31 $\pm$ 1.12 & 74.10 $\pm$ 0.73             \\
BCI-COMP               & \boldmath$86.39 \pm 1.13$                   & \boldmath$80.11 \pm 6.35$  & $71.61 \pm 2.97$ &     $77.20 \pm 4.18$ & $72.15 \pm 3.01$ &     $77.56 \pm 1.73$   \\
Auditory-BCI                         & 83.43 $\pm$ 4.52                   & 50.00 $\pm$ 0.00   & 83.43 $\pm$ 2.62                   & 50.12 $\pm$ 0.07      & 83.47 $\pm$ 3.17 & 64.69 $\pm$ 0.21           \\
BCI-Spell                     & 83.33 $\pm$ 4.62                   & 50.00 $\pm$ 0.00   & 83.58 $\pm$ 5.93                   & 49.56 $\pm$ 0.53  & 83.51 $\pm$ 2.31                   & 49.22 $\pm$ 0.79      \\
EEG-Alcochol  & \boldmath$79.37 \pm 2.41$                   & \boldmath$87.39 \pm 1.42$  & \boldmath$79.84 \pm 2.46$& \boldmath $86.52 \pm 1.92$   & 70.95 $\pm$ 3.32 &86.26 $\pm$ 1.12        \\
Decode-Audi     & 54.68 $\pm$ 1.83     & 52.52 $\pm$ 0.03  & 51.93 $\pm$ 0.21 & 50.92  $\pm$ 0.05   & 46.44 $\pm$ 1.51 & 49.07  $\pm$ 0.95         \\ \hline
\end{tabular}

\begin{tabular}{lllll}
\\
\hline
\textbf{}                                       & \multicolumn{2}{c}{\textbf{SSLC-AE}}         & \multicolumn{2}{c}{\textbf{XdawnLDA}}\\ \hline
\multicolumn{1}{c}{\textbf{dataset}} & \multicolumn{1}{c}{\textbf{ACC}} & \multicolumn{1}{c}{\textbf{AUC}} & 
\multicolumn{1}{c}{\textbf{ACC}} & \multicolumn{1}{c}{\textbf{AUC}} \\ \hline
P300-Amplitude  &  $83.43 \pm1.87$ & $78.29\pm 1.14$            & 76.25 $\pm$ 0.71                   & 77.22 $\pm$ 2.11 \\
BCI-COMP & $68.66\pm3.53$ & $74.58\pm 0.85$                  & 63.17 $\pm$ 0.01     & 73.33 $\pm$ 2.85 \\
Auditory-BCI  & $83.43\pm3.53$ &   $52.86\pm 2.92$           & 60.56 $\pm$ 1.28          & 61.16 $\pm$ 10.5\\
BCI-Spell  & $83.58\pm4.72$ & $49.26\pm1.97$ & 83.33 $\pm$ 4.62  & 50.00 $\pm$ 0.00         \\
EEG-Alcochol & $76.56\pm2.36$ & $83.83\pm1.63$             & 75.79 $\pm$ 0.75   & 83.28 $\pm$ 0.85\\
Decode-Audi  & $51.43\pm4.26$ & $50.36\pm1.63$       & 50.10 $\pm$ 0.31      & 50.16 $\pm$ 0.31     \\ \hline
\end{tabular}

\end{table*}
Table \ref{resulttable2} summarizes the quality of the latent vector by comparing the ACC and AUC. Statistically, we cannot reject the null hypothesis that the ACC and AUC are equal to the Wilcoxon signed-rank test in every dataset. Only in one dataset, namely P300-Amplitude, ERPENet outperforms the SSLC-AE, while SSLC-AE outperforms the ERPENet in two datasets. SSLC-AE performs slightly better than our proposed model. However, as we will show in Experiment B, fine-tuning was required for efficient use of the ERPENet. This is consistent with a recent study in computer vision that better models do not necessary extract better features without appropriate adaptation of the feature extraction network \cite{kornblith2018better}.

\subsection{Experiment B: Adaptation of pre-trained model(ERPENet)}
\begin{figure}[H]
  \centering
  \scalebox{0.6}{\input{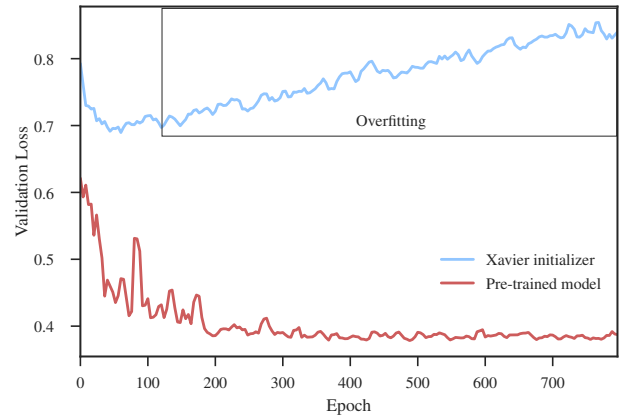}}
  \caption{The validation loss from training the ERPENet model on BCI-Spell dataset using pre-trained weights and Xavier initialization.}
  \label{fig:loss}
\end{figure}

 In the Xavier weight initialized trainings, the validation losses show signs of overfitting around epoch 25 on the BCI-COMP dataset and around epoch 100 on BCI-Spell dataset, as shown in Figure \ref{fig:loss}. This signifies the important of training ERPENet with joint datasets. Moreover, the validation loss value indicates that the pre-trained model at epoch 0 already outperforms all epochs of the Xavier initialized model.

In Table \ref{resulttable3}, there are only three datasets (P300-Amplitude, BCI-COMP, EEG-Alcohol) that can be discriminated well by all classifiers (ERPENet, ERPENet, DeepConvNet, SSLC-AE and XdawnLDA), indicated by the AUC higher than 70. The Wilcoxon signed-rank test, a two-sided p-value of null hypothesis, is computed to compare the performance between ERPENet and four baseline models. The test yields a mean difference in significant values less than 0.01 for all three datasets. The result indicates that the ERPENet outperformed all models in the three datasets, except for EEGNet tested by EEG-Alcohol dataset. Moreover, adaption of the ERPENet in the proposed multi-task AE improves performance both in time efficiency and classification accuracy showing the importance of adaptation.

\section{Discussion}
In Experiment A, the SSLC-AE is much shallower than our proposed model, converging in only 295 epochs. On the other hand, our proposed multi-task AE was trained for 833 epochs. The training epochs required to train our model were about three times greater, representing a trade-off for a more complex model.

In Experiment B, EEGNet and DeepConvNet were pre-trained in the same way as ERPENet. Regardless of the compact size in EEGnet and the claim in \cite{Waytowich2018} that EEGNet could be trained with very limited data, EEGNet did not perform well in BCI-COMP which is the smallest dataset in this evaluation. Table \ref{resulttable3} also shows that ERPENet outperforms DeepConvNet, while DeepConvNet has comparable results as in EEGNet, which is consistent with the experiments with P300 dataset in \cite{Waytowich2018}. ERPENet has a larger size of trainable parameters and longer training and inference times than the others, but with the speed of nowadays GPU, ERPENet is capable of predicting more than 2600 P300 trials per second.

Among all datasets, BCI-COMP, EEG-Alcohol and P300-Amplitude have the smallest number of samples, with ERPENet obtaining a higher ACC and AUC than the Xdawn algorithm on all three datasets. It could be inferred that the pre-trained model would improve the training of DL models on datasets with a small number of samples.

In ERPENet training, multiple P300 datasets from various sources and tasks were combined together to improve the model. There was some incompatibility in recording standards and protocols across the datasets, increasing the bias of the model. Before incorporating the supervised classifier part, experiments were also conducted using variational AE, but high diversity between P300 tasks prevented the models from reaching the optimal points and ultimately overfited to the dominated dataset. 

In this work, comparisons across all datasets were possible, since the EEG was normalized before the training process. From Table \ref{resulttable1}, Auditory-BCI, BCI-Spell, and Decode-Audi datasets had higher MSE compared to the others for both the multi-task AE and SSLC-AE. In the classification tasks, the AUCs of auditory-BCI, BCI-Spell, and Decode-Audi datasets were below 70 for all methods. A single FC in ERPENet was considered a simple model which might not be suitable classifiers for these three tasks. Increasing the size of the classifier without pre-trained weight might lead to overfitting problem which we would like to avoid. Additionally, Auditory-BCI and Decode-Audi are both auditory tasks, and P300 might not be able to capture all of the critical features. Other constituents of the ERP, such as N200 and increasing the size of the classifier network could be considered for inclusion in future works.

\section{Conclusion}
In conclusion, we have shown that our proposed multi-task AE, incorporating CNN and LSTM, has the capability for better compression than the previously proposed multi-task AE (SSLIC) composed of FCs, while maintaining high accuracy in the prediction of attended and unattended events on single trials of P300 EEG. Moreover, the encoder part of our proposed model can be extended as a pre-trained network, namely ERPENet, for other P300 tasks thereby reducing overfitting during training and hastening the training of complex models. This extended classification model also outperformed EEGNet and DeepConvNet, which are state-of-the-art deep learning models in EEG classification tasks, along with a state-of-the-art dimensional reduction algorithm designed for P300, Xdawn. This is a pioneer work that proposes the concept of the pre-trained networks for other EEG-related applications.

\bibliographystyle{IEEEtran}
\bibliography{ref2}

\begin{IEEEbiography}[{\includegraphics[width=1.1in,height=1.7in,clip,keepaspectratio]{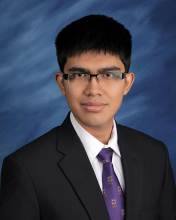}}]
{Apiwat Ditthapron} is currently pursuing B.S. degree from the department of computer science, Worcester Polytechnic Institute, MA, USA. His current research interests include Computer Vision, Machine learning, Deep learning, and Data Visualization.
\end{IEEEbiography}

\vskip -2pt plus -1fil

\begin{IEEEbiography}[{\includegraphics[width=1in,height=1.25in,clip,keepaspectratio]{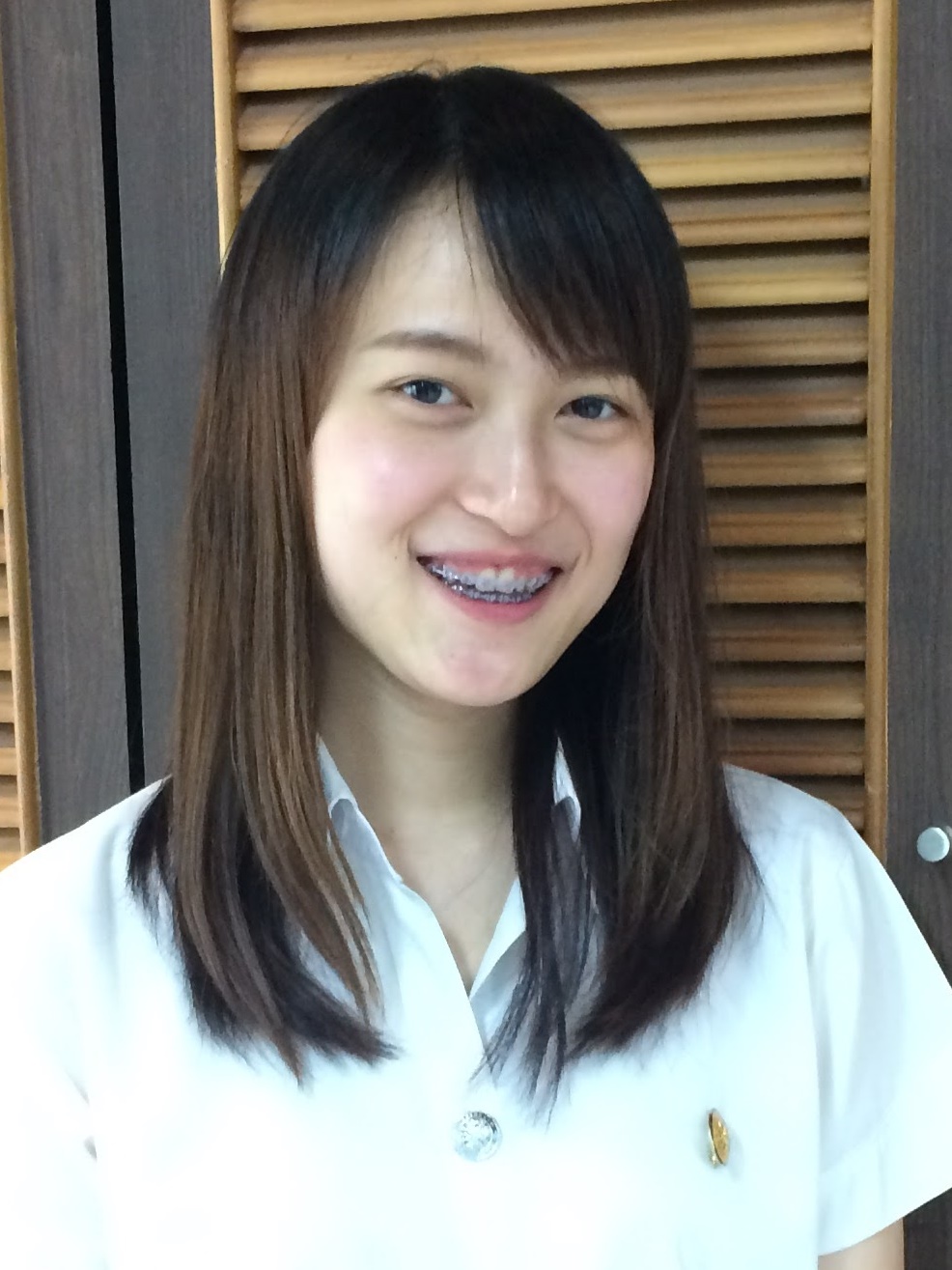}}]
{Nannapas Banluesombatkul} recieved the B.Sc. degree in Computer Science from Thammasat University, Thailand in 2017. She is currently a Research Assistant with Bio-inspired Robotics and Neural engineering (BRAIN) lab, School of Information Science and Technology at Vidyasirimedhi Institute of Science and Technology (VISTEC), Thailand. Her current research interests include biomedical signal processing and clinical diagnosis support system.
\end{IEEEbiography}
\vskip -2pt plus -1fil

\begin{IEEEbiography}[{\includegraphics[width=1.1in,height=1.6in,clip,keepaspectratio]{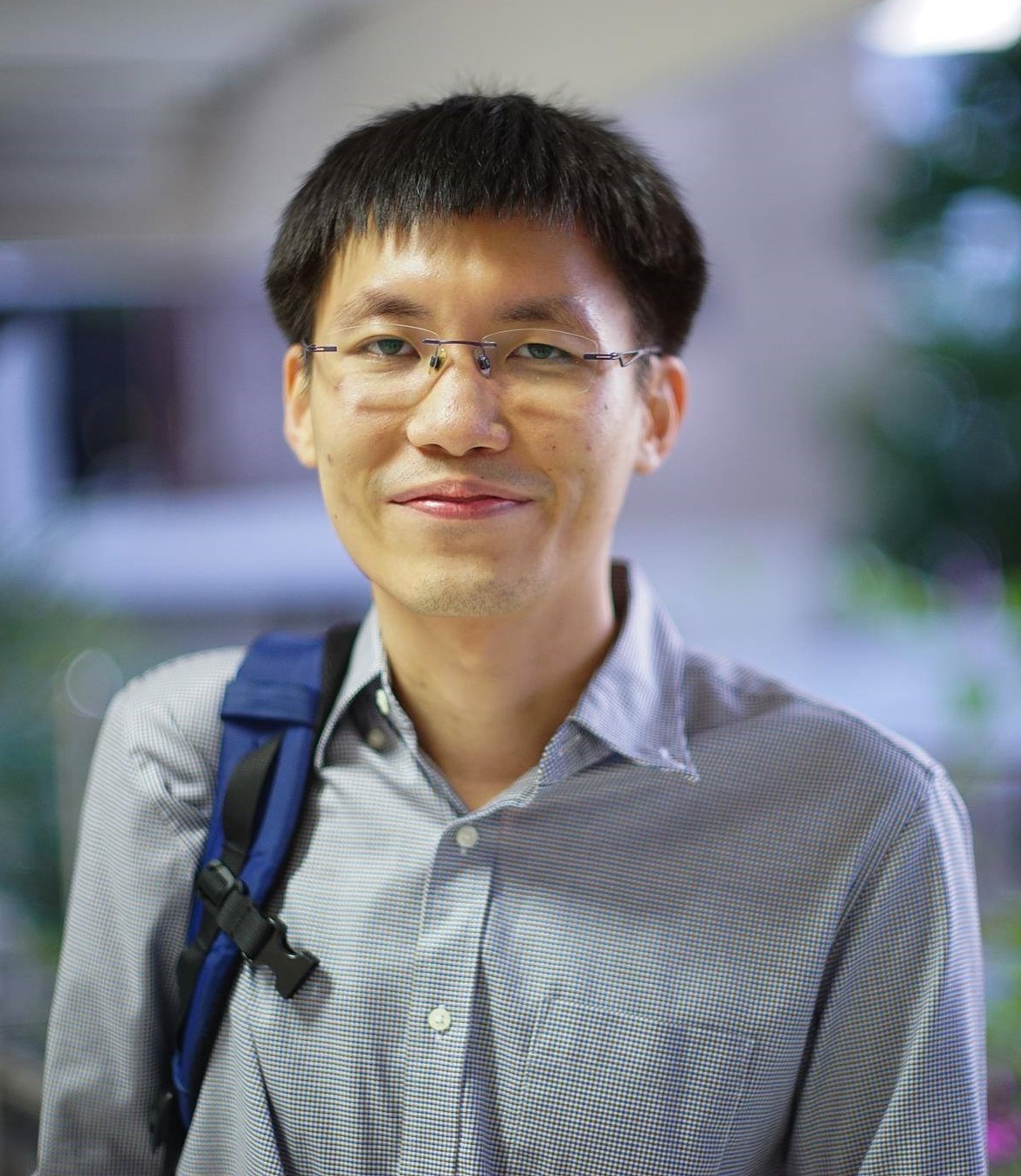}}]
{Ekapol Chuangsuwanich} received the B.S. and S.M. degree in Electrical and Computer Engineering from Carnegie Mellon University in 2008 and 2009. He then joined the Spoken Language Systems Group at MIT Computer Science and Artificial Intelligence Laboratory. He received his Ph.D. degree in 2016 from MIT. He is currently a Faculty Member of the Department of Computer Engineering at Chulalongkorn University. His research interests include machine learning approaches applied to speech processing, assistive technology, and health applications.
\end{IEEEbiography}

\vskip -2pt plus -1fil
\begin{IEEEbiography}[{\includegraphics[width=1.1in,height=1.4in,clip,keepaspectratio]{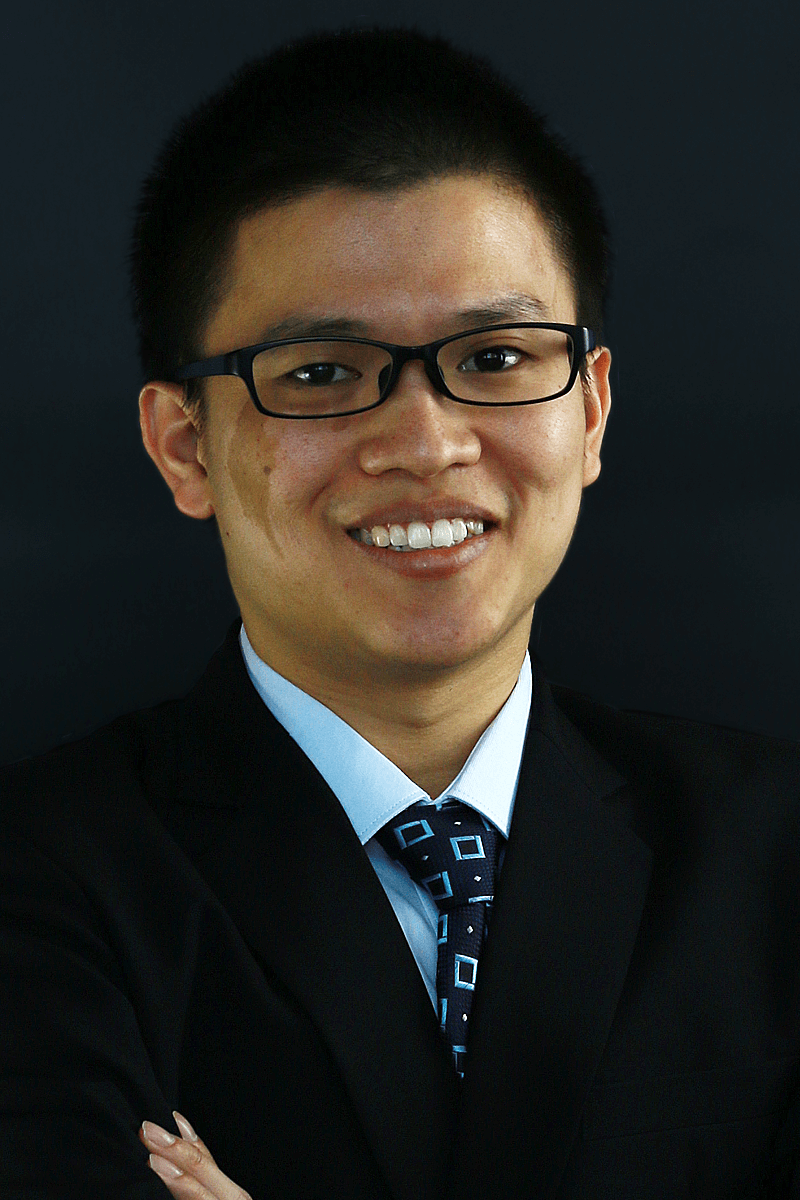}}]{Theerawit Wiaiprasitporn}
received the Ph.D. Degree in Engineering from Graduate School of Information Science and Engineering, Tokyo Institute of Technology, Japan, in 2017. As a Ph.D. student, he obtained own research funding from Japan Society for the Promotion of Science (JSPS). While pursuing his Degree, he did short-term research at NASA Ames Research Center in USA . Now, he is working as lecturer position at School of Information Science and Technology at Vidyasirimedhi Institute of Science and Technology (VISTEC), Thailand. His current research are Neural Engineering (BCI), Bio-Potential Applications, Biomedical and Health Informatics and Smart Living.
\end{IEEEbiography}

\EOD

\end{document}